\documentclass[%
 reprint,
 amsmath,amssymb,
 aps,
]{revtex4-1}

\usepackage{graphicx}
\usepackage{dcolumn}
\usepackage{bm}
\usepackage{epstopdf}
\usepackage{amsmath}
\usepackage{lipsum} 
\usepackage{mathtools}
\usepackage{caption}
\usepackage{subfigure}
\usepackage{textcomp}
\usepackage{upgreek}
\usepackage{float}
\usepackage{ragged2e}
\usepackage{caption}

\begin{document}


\title{Artificial Intelligence Enhances the Performance of Chaos-based Wireless Communication}

\author{Hai-Peng Ren}
\author{Hong-Er Zhao}%
\author{Chao Bai}
\author{Hui-Ping Yin}
\affiliation{%
 Shaanxi Key Laboratory of Complex System Control and Intelligent Information Processing, Xi’an University of Technology, Xi’an 710048, China\\
 (Corresponding author e-mail:renhaipeng@xaut.edu.cn)
}%
\author{Celso Grebogi}
\affiliation{%
 Institute for Complex Systems and Mathematical Biology, University of Aberdeen AB24 3UE, United Kingdom.
}%


\begin{abstract}
Some new findings for chaos-based wireless communication systems have been identified recently. First, chaos has proven to be the optimal communication waveform because chaotic signals can achieve the maximum signal to noise ratio at receiver with the simplest matched filter. Second, the information transmitted in chaotic signals is not modified by the multipath wireless channel. Third, chaos properties can be used to relief inter-symbol interference (ISI) caused by multipath propagation. Although recent work has reported the method of obtaining the optimal threshold to eliminate the ISI in chaos-based wireless communication, its practical implementation is still a challenge. By knowing the channel parameters and all symbols, especially the future symbol to be transmitted in advance, it is almost an impossible task in the practical communication systems. Owning to Artificial intelligence (AI) recent developments, Convolutional Neural Network (CNN) with deep learning structure is being proposed to predict future symbols based on the received signal, so as to further reduce ISI and obtain better bit error rate (BER) performance as compared to that used the existing sub-optimal threshold. The feature of the method involves predicting the future symbol and obtaining a better threshold suitable for time variant channel. Numerical simulation and experimental results validate our theory and the superiority of the proposed method.
\begin{description}
\item[PACS numbers]
0545
\end{description}
\end{abstract}

\pacs{0545Valid PACS appear here}
\maketitle


\section{\label{sec:level1}Introduction}
Artificial intelligence (AI) is a new technology for simulating and emulating human cognition. It was proposed in 1956 and has experienced a tortuous development [1,2]. Now AI has been demonstrating its superior performance in a variety of practical applications [3]. Convolutional Neural Network (CNN) is one of the most popular deep learning network structures and has a wide range of applications in various fields. In the field of communication, the increase of hardware computing speed, especially in the field-programmable gate array (FPGA) with high performance to price ratio, makes it possible to apply AI in communication systems [4-7].

Chaos has been applied in communication since 1993 [8,9], but most early research concentrated on the theoretical analysis of the bit error rate and security in the ideal channel, possibly with white noise. Since reference [8] reported performance improvement of the fiber-optical communication using chaos, more attention has been attracted into practical communication channel applications. The wireless channel, the topic of this work, is a practical channel widely used. Due to the serious distortion of the wireless communication channel caused by limited bandwidth, multipath propagation, and complicated noise, etc., the wireless communication requires more sophisticated technique to achieve satisfactory performance as compared to the wired communication counterpart. Although chaotic signals have been reported to possess some properties fit for communication, such as broad band, orthogonality, etc., whether the information in the chaotic signal suffers loss was a pending problem until it has been shown that the Lyapunov exponent spectrum of the chaotic signal after transmitted through wireless channel is unaltered, which means that the information content in it is not lost [9,10]. Further efforts have boosted the chaos-based wireless communication systems after some new features due to chaos were reported. They include that the chaotic signal has a very simple corresponding matched filter to maximize the signal to noise ratio, the invariant Lyapunov exponent spectrum can be used to relieve the inter-symbol interference (ISI) caused by multipath propagation [11].

Recently, ISI resistance performance of chaos-based wireless communication systems was theoretically and experimentally investigated [11-13], where using chaos properties to relieve ISI was shown to result in superior performance, as compared to the conventional method, i.e., channel equalization. The current ISI relief technique in [11-13] only employs the available information, i.e., the past received bits (symbols), because the future symbol is not available at the current time, even though it affects the ISI relief performance. How to improve the ISI relief performance by predicting or estimating the future information bit? This has become a critical issue, which we address in this \emph{Letter}. Thanks to recent developments in AI, powerful tools to predict future time series [14-17] is becoming now available. Particularly, in our case, the final goal is to predict the future information symbol sequence from the time series. Here, we employ the CNN to infer the symbol directly, in such a way avoiding the intermediate time sequence prediction and simplifying the operation. Besides this advantage, in our wireless communication system configuration, the data is transmitted frame by frame with probe data for these purposes, such as, clock synchronization. The probe data is used for CNN training to avoid the time varying parameters effect on the ISI relief performance.
\section{\label{sec:level1}Method}
CNN was first put forward in 1962 by Hubel and Wiesel in connection with the cat visual cortex, which effectively reduces the learning complexity [18]. Based on the result in [19], Yann LeCun et al. designed an error gradient-based algorithm to train CNN, which showed superior performance in some pattern recognitions as compared to other methods [20]. This result led to growing interest in the research on CNN [21-24].

\begin{figure}[h]
\centering
\includegraphics[width=2.5in]{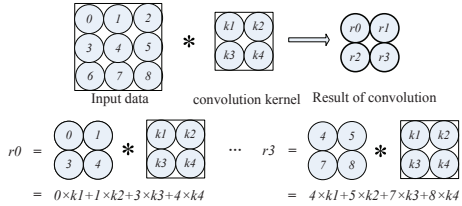}
\captionsetup{font={small}}\caption{Convolution of a $3\times3$ input data with a $2\times2$ convolution kernel, where $*$ represents convolution.}
\label{fig:arch} 
\end{figure}

CNN consists of three basic layers: convolutional layer, pooling layer and full connection (output) layer. A convolution layer is used to extract the feature of the input data. Using a $3\times3$ input data as an example, if the convolution kernel size is $2\times2$, the weight of the convolution kernel is $k1$, $k2$, $k3$, $k4$. Then the convolution process using the given convolution kernel and input data is pictorially shown in Fig. \ref{fig:arch}.

As seen from Fig. 1, for this $3\times3$ input data, we use a $2\times2$ convolution kernel to convolve. By setting the slide step size as 1, that is, sliding the $2\times2$ window to the right or down by one element at a time, the convolution result is obtained. Different convolution kernels extract the different features of the input data, so the convolutional layer function is the feature extraction.

The purpose of the pooling layer is to compress data by downsampling. Average pooling is used here. In the average pooling method, the elements of a selected size window are used to calculate an average value. Using a $2\times2$ window [r0, r1; r2, r3] as example, after the average pooling, the output data from this $2\times2$ window is obtained as follows:
\begin{equation}
p0={\left( r0+r1+r2+r3 \right)}/{4}\;.
\end{equation}
For all pooling layer input data, sliding the window with the step size, we get all output of the pooling layer. The pooling layer is usually placed between two convolutional layers, which acts as the secondary feature extraction, in order to reduce the risk of over-fitting, to reduce the dimension of the feature map and to improve the robustness of feature extraction. The high level features can be extracted by stacking the convolutional layers and pooling layers several times.

The fully-connected layer has all to all connection with neurons in the previous layer. This layer is followed by the output of the network. Then the regression function is used here to complete the classification task. This part acts as a classifier, which is referred to the fully-connected layers with respect to the previous layer [25]. The output calculation formula is given as follows:
\begin{equation}
\begin{small}
\left\{ \begin{gathered}
{{y}_{1}} ={{b}_{1}}\times 1+{{w}_{11}}\times {{x}_{1}}+{{w}_{21}}\times {{x}_{2}}+\cdots+{{w}_{n1}}\times {{x}_{n}} \\
\cdots \\
{{y}_{m}} ={{b}_{m}}\times 1+{{w}_{1m}}\times {{x}_{1}}+{{w}_{2m}}\times {{x}_{2}}+\cdots+{{w}_{nm}}\times {{x}_{n}} \\
\end{gathered} \right.,
\end{small}
\end{equation}
where ${{y}_{j}}$ ($j=1, \cdots, m$) is the output, ${{x}_{i}}$ ($i=1, \cdots, n$) is the neuron in the layer before the output layer, ${{b}_{j}}$ is the bias of ${{y}_{j}}$ and ${{w}_{ij}}$ is the weight from ${{x}_{i}}$ to ${{y}_{j}}$.

The training process of CNN consists of two stages. The first stage is a feedforward calculation process, where data are transmitted from input to output layer via intermediate layers. The second stage is the error back propagation from the output layer to the input layer when the current output is not equal to the expectation, that is, back propagation [26]. When the CNN output does not match the expectation, the error between the current output and the expected value is propagated layer by layer back in order to calculate the error contribution of each layer, and then the weights of each layer are updated accordingly. After updating the weights, the input and the new weights are used to calculate the new output. If the new output matches the expectation, the training is completed, and if not, the error back propagation and weight updating are repeated again until the output match.

As described in Refs. [11-12], the ISI in chaos-based wireless communication system is given by:
\begin{equation}
\begin{small}
\begin{aligned}
  {{\theta }_{n}} &={{e}^{-\frac{\beta }{f}}}{{y}_{n+1}}+\sum\limits_{l=0}^{L-1}{\sum\limits_{\begin{smallmatrix}
 i=-\infty  \\
 i\ne 0
\end{smallmatrix}}^{1}{{{s}_{n+i}}\left( {{C}_{l,i}}-{{e}^{-\frac{\beta }{f}}}{{C}_{l,i-1}} \right)}} \\
 &={{I}_{past}}+{{I}_{future}}=I,
\end{aligned}
\end{small}
\end{equation}
where ${{s}_{n+i}}\left( i<0 \right)$ is the past transmitted symbols and ${{s}_{n+i}}\left( i>0 \right)$ is the future transmitting symbols. ${{\theta }_{n}}$ is the optimal threshold to decode the received symbol, which is the sum of ${{I}_{past}}$ and ${{I}_{future}}$, where ${{I}_{past}}=\sum\limits_{l=0}^{L-1}{\sum\limits_{i=-\infty }^{i=-1}{{{s}_{n+i}}{{C}_{l,i}},}}{{s}_{n+i}}\left( i<0 \right)$ is the ISI caused by past symbols and   ${{I}_{future}}=\sum\limits_{l=0}^{L-1}{\sum\limits_{i=1}^{i=\infty }{{{s}_{n+i}}{{C}_{l,i}},}}{{s}_{n+i}}\left( i>0 \right)$ is the ISI caused by future symbols. ${{C}_{l,i}}$ can be calculated as following, ${{\alpha }_{l}}$ and ${{\tau }_{l}}$ being the channel parameters obtained by channel parameters estimation [27]:
\begin{equation}
\begin{small}
{{C}_{l,i}}=\left\{ \begin{aligned}
  &{{c}_{1}}, &for\left( \left| {{\tau }_{l}}+{i}/{f}\; \right|\ge {1}/{f}\; \right) \\
 \ &{{c}_{2}}, &for\left( 0\le \left| {{\tau }_{l}}+{i}/{f}\; \right|<{1}/{f}\; \right) \\
 \end{aligned} \right.,
\end{small}
\end{equation}
where ${{c}_{1}}$ and ${{c}_{2}}$ are defined in [28]. It can be seen from Eqs. (3) and (4) that the optimal threshold is determined by three factors, namely the past received information symbols (bits), the future (sending) symbols, which is not received yet at current time, and the channel parameters needed in Eq. (4). The past received symbols, ${{s}_{i}}\left( i<0 \right)$, have been decoded, which is available at current time. However, the future information symbols, ${{s}_{i}}\left( i>0 \right)$, which have not been received yet and undetermined at the current time. Therefore, the current technique, using the threshold determined by the past information symbols, i.e., ${{I}_{past}}$, decodes the information bit to achieve sub-optimal performance as compared to one the using optimal threshold, i.e., the sum of ${{I}_{future}}$ and ${{I}_{past}}$. Although using ${{I}_{past}}$, it still achieves significantly better performance as compared to the conventional method, which uses 0 as threshold and channel equalization [29,30], it does leave space for further improvement.

\begin{figure}[h]
\centering
\includegraphics[width=3.1in]{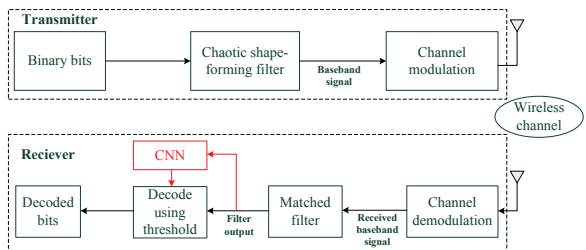}
\captionsetup{font={small}}\caption{Block diagram of chaotic baseband wireless communication system.}
\label{fig:arch5} 
\end{figure}

We thus further improve the performance, using CNN in this \emph{Letter} to predict the future symbols to be sent. In this way, the more accurate decoding threshold is obtained and used to improve the BER performance. The block diagram of our chaos baseband wireless communication system is given in Fig. \ref{fig:arch5}.

In chaos-based wireless communication system [31], the chaotic baseband signal is generated by chaotic shape-performing filter given by the following equation.
\begin{equation}
u\left( t \right)=\sum\nolimits_{m=-\infty }^{\infty }{{{s}_{m}}\cdot }p\left( t-m \right),
\end{equation}
where $p\left( t \right)$ is the basis function:
\begin{equation}
p\left( t \right)=\left\{ \begin{aligned}
  & \left( 1-{{e}^{-\beta }} \right){{e}^{\beta t}}\left( \cos \omega t-\left( {\beta }/{\omega }\; \right)\sin \omega t \right),&\left[ t \right]<0 \\
 & 1-{{e}^{-\beta \left( t-1 \right)}}\left( \cos \omega t-\left( {\beta }/{\omega }\; \right)\sin \omega t \right),&\left[ t \right]=0 \\
 & 0,&\left[ t \right]>0 \\
\end{aligned} \right.,
\end{equation}
$\omega =2\pi f$ is the base frequency, $\beta =f\ln 2$, ${{s}_{m}}\in \left[ -1,1 \right]$ is the information symbol to be transmitted.

In Fig. \ref{fig:arch5}, the binary information bits to be sent are fed into the chaotic shape forming filter given by Eq. (5) to form the baseband signal, then the baseband signal is passed through the channel modulation to get the signal transmitted in the physical wireless channel. After the wireless channel transmission, at the receiver, the received signal undergoes the channel demodulation to remove the channel modulation frequency and derive the received baseband signal, which passes through the matched filter and is sampled at the same over sampling rate as that of the shape-forming filter. In this work, the over sampling rate is ${{n}_{samp}}=16$.

\begin{figure}[h]
\centering
\includegraphics[width=2.6in]{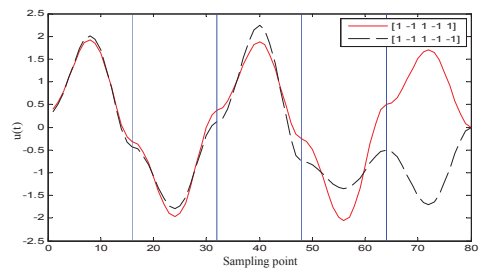}
\captionsetup{font={small}}\caption{The difference in the past baseband signal caused by the future one bit prediction.}
\label{fig:arch6} 
\end{figure}

The baseband signal of the current information symbol is affected by both the past information symbols and the future ones. This fact makes it possible for us to use the (past) received signal to predict the future symbol. In Fig. \ref{fig:arch6}, the past 4 symbols are the same, i.e., [1, -1, 1, -1], however, the difference in the past baseband signal caused by future one bit (1 or -1) prediction can be observed.

From Fig. \ref{fig:arch6}, we draw two conclusions: first, importantly, the future symbol (bit) difference will cause the current waveform difference, which is perceivable by the proper feature collection using CNN; second, the further is the future symbol, the less influence has on the past waveform, which could also be obtained from Eq. (6). These two points are very important for our work. First, it is possible for us to use current waveform to predict future symbol because the different future symbol causes the difference in current waveform. Second, it is easy for us to use not very long past waveform to predict future symbol because the further future symbol causes less effect on the past waveform.

\begin{figure}[H]
\centering
\includegraphics[width=2.1in]{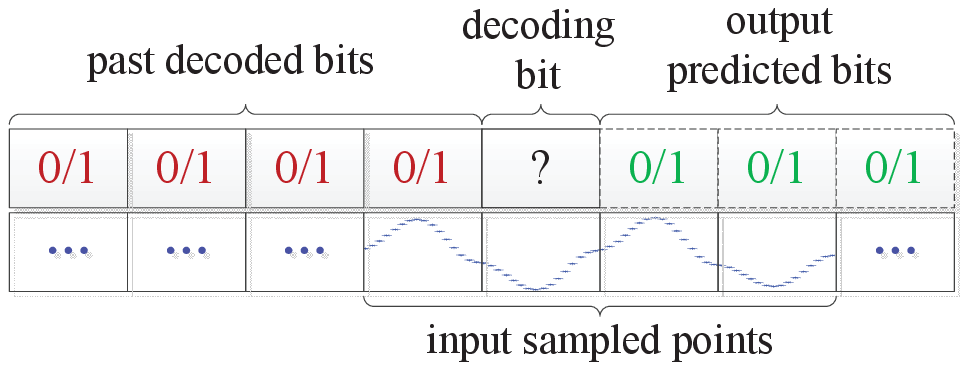}
\captionsetup{font={small}}\caption{The input and output of CNN.}
\label{fig:arch9} 
\end{figure}

In our chaos-based wireless communication system, the data is transmitted frame by frame so that the channel parameters can be treated as time-invariant parameters. One data frame consists of the probe data and the information data. The probe data are the known signals generated by our train data passing shape forming filter instead of conventional logistic map, which is used for clock synchronization, frequency bias compensation, the channel parameter identification, as well as the training data in our scheme. In our configuration, from Fig. \ref{fig:arch6} and the description in the last paragraph, we are not able to predict very far future symbol, because the further future data, the less effect can have on the current waveform. Here we select the waveform corresponding to the past four data, which means 64 sampled points, being $8*8$ matrix input data to CNN. The expected output is a vector consisting of one future bit, the current bit and one past bit. As seen from Fig. \ref{fig:arch9}, in the training stage, the sampled points of 4 symbols (with over sampling rate ${{n}_{samp}}=16$), which is a $8\times8$ matrix, are used as CNN input, and the CNN expected output is the binary bits as marked ``output predicted bits" in Fig. \ref{fig:arch9}. In the prediction (application) stage, the input and output are the same as that in the training stage, then the 4 past decoded bits and the 3 predicted future bits are used together in Eq. (3) to calculate more accurate decoding thresholds for bit marked by ``?" in Fig. \ref{fig:arch9}. In the CNN training state, four past symbols plus one future bit would possess ${{2}^{6}}$ kinds of possible value group, which corresponds to ${{2}^{6}}*8*8$ matrix input training data with expected output from 0 to 7 correspondingly, when the output vector is treated as a 3 bit binary number.
\noindent One thing to notice here is that the channel parameters are not input to the CNN, because the channel parameters are assumed to be constant in one frame of transmitting data, and the channel information has been encoded into the received waveform, which simplifies the predicting operation.

\section{\label{sec:level1}Simulation Results}
In the simulation, the wireless channel is modeled by
$h\left( t \right)=\sum\nolimits_{l=0}^{L-1}{{{\alpha }_{l}}\delta \left( t-{{\tau }_{l}} \right)},$
where ${{\alpha }_{l}}={{e}^{-\gamma {{\tau }_{l}}}}$ is the attenuation of path $l$, ${{\tau }_{l}}$ is time delay of path $l$. $L$ is the multipath number. $\delta \left( \cdot  \right)$ is the Dirac delta function [12,31]. Here the base frequency is $f=600Hz$.

Figure \ref{fig:subfig:}(a) shows BER for the different methods in the case of single path versus BitEnergy-to-NoiseDensity (Eb/N0). $\theta=0$ indicates the BER curve that does not consider ISI, $\theta=Ipast$ indicates the BER curve that only considers ISI caused by past 4 symbols, as done in [11, 12], and the proposed method considers the ISI caused by both past 4 symbols and future 3 symbols. It can be seen from Fig. \ref{fig:subfig:}(a) that the proposed method has the lowest BER, so it shows that this method can reduce the ISI in the case of single path.

Figure \ref{fig:subfig:}(b) shows BER for the different methods versus Eb/N0 for three-path channel, whose parameters are given in [32]. It can be seen from Fig. \ref{fig:subfig:}(b) that the proposed method has the lowest BER. We can also see that the decoding without considering ISI becomes even worse as compared to the single path case, since in this case the ISI becomes stronger as compared to the single path case.

\section{\label{sec:level1}Experimental Results}
In the experiment, two Wireless Open-Access Research Platform (WARP) with Virtex-6 LX240T FPGA are used [33]. Here ${{n}_{samp}}=16$, baseband frequency $f=40 M Hz$, and the carrier frequency is $2.4 G Hz$.

\begin{figure}[H]
\centering
\includegraphics[width=3.5in]{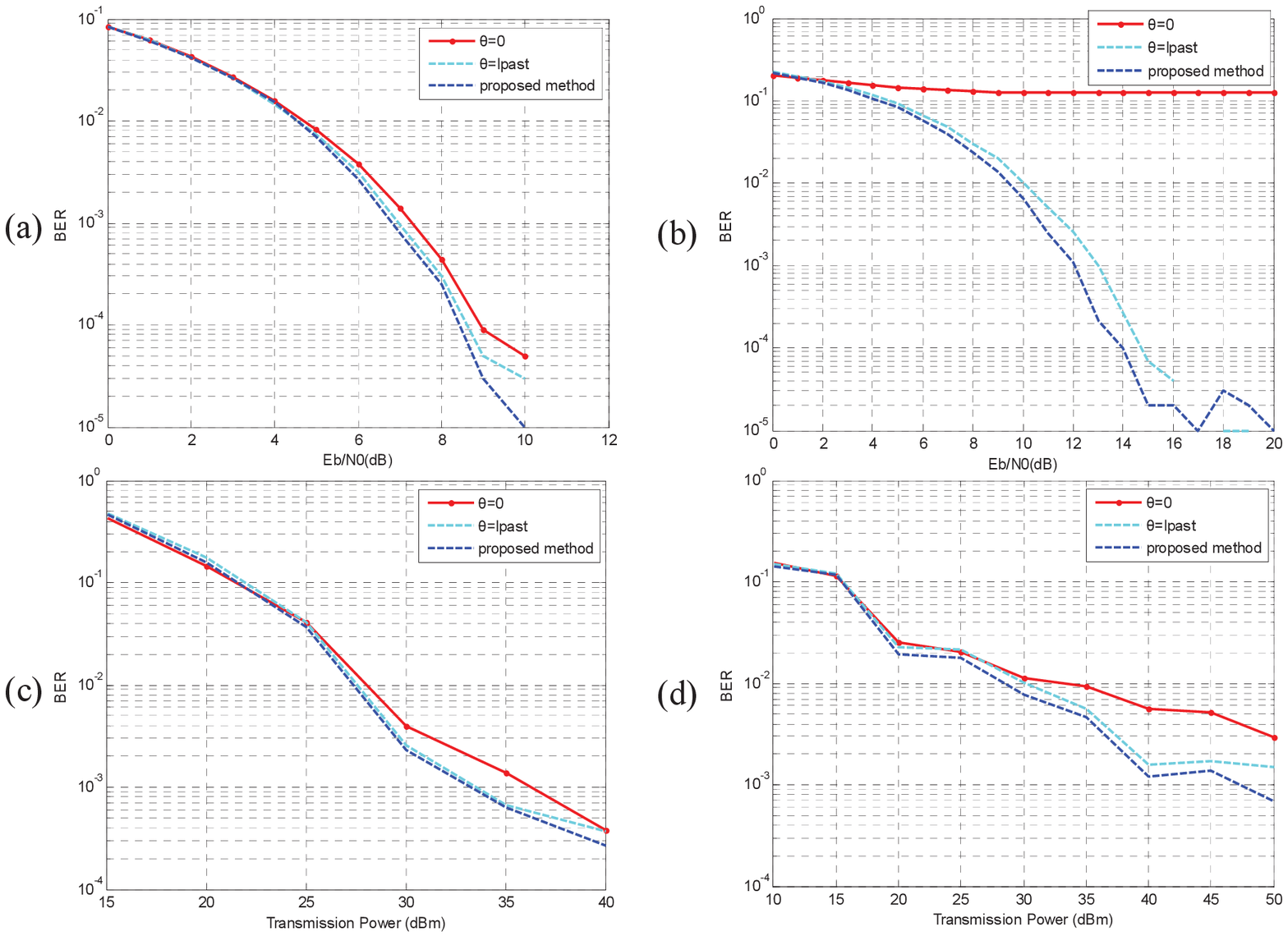}
\captionsetup{font={small}}\caption{BER comparison for simulation over single path (a) and multipath (b), using parameters in [32], as well as for experiment over single path (c) and multipath (d), with parameters in [34].}
\label{fig:subfig:} 
\end{figure}

First, the transmitter and receiver are placed in the laboratory to test the single path case.  Figure \ref{fig:subfig:}(c) shows BER for the different methods versus the transmission power. In WARP, the transmission power can be adjusted by parameters TX\_RF\_Gain and TX\_BB\_Gain to simulate the signal to noise ratio variation. It can be seen from Fig. \ref{fig:subfig:}(c) that the proposed method has the lowest BER, showing that this method reduces the ISI in the case of single path channel (with noise) by predicting future symbols.

We also tested experimentally the BER performance under the real multipath channel. In this scenario, the distance between TX and RX is about 15 meters. We have verified that the channel does not vary during the transmission of one frame in our test [12]. Figure \ref{fig:subfig:}(d) shows BER for the different methods versus the different transmission power. It can be seen that the proposed method has the lowest BER also in the case of the multipath channel.
%

\section{\label{sec:level1}Conclusion}
In this \emph{Letter}, a method based on CNN is being proposed to reduce the ISI in chaos-based wireless communication system and decrease BER. The previous method can only eliminate the ISI caused by the past symbols because the future information symbols are unavailable. In order to further decrease the ISI caused by the future information symbols and enhance the performance of chaos-based wireless communication, we use AI to predict future information symbols. By training, the CNN has the ability to recognize the feature of the input data, and then predict the future symbol to be received based on the waveform of past symbols. Other tests also proves that, under different delay and attenuation conditions, a lower BER can be achieved by using the proposed method, which are not given here because of a limited space. The experimental tests pave the way for practical application of chaos-based wireless communication with the hardware compatible with the traditional system but higher performance is achieved by updating the software.

This work is supported by National Natural Science Foundation of China (61172070) and Shaanxi Provincial Special Support Program for Science and Technology Innovation Leader.


\begin{thebibliography}{31}
\bibitem[32]{ref32}In our simulation of three-path channel, ${{\tau }_{0}}$= 0 ms, ${{\tau }_{1}}$= 26.7 ms, ${{\tau }_{2}}$= 53.3 ms, ${{\gamma }}$= 0.4, thus ${{\alpha }_{1}}$= 0.6703, ${{\alpha }_{2}}$= 0.4493.
\bibitem[34]{ref34}There are four main paths identified by channel estimation with channel fading 1, 0.28, 0.16 and 0.18, respectively, and the delay of four paths are 0 $\upmu$s, 0.4 $\upmu$s, 0.8 $\upmu$s and 1.2 $\upmu$s, respectively.
\end{thebibliography}
\end{document}